\documentclass[preprint,aps,psfig,epsfig,showkeys,showpacs]{revtex4}

\begin{document}

\title{Bose-Einstein Condensation in Random Directed Networks}
\author{O. Sotolongo-Costa$^{1,3}$ and G. J. Rodgers$^{2,3}$}
\affiliation{$^{1}$Department of Theoretical Physics, Havana University, 10400 Havana,Cuba}
\affiliation{$^{2}$Department of Mathematical Sciences, Brunel University, Uxbridge,
Middlesex UB8 3PH, U.K.}
\affiliation{$^{3}$ Henri Poincare Chair of Complex Systems, Havana University.}

\begin{abstract}
We consider the phenomenon of Bose-Einstein condensation in a random growing
directed network. The network grows by the addition of vertices and edges.
At each time step the network gains a vertex with probabilty $p$ and an edge
with probability $1-p$. The new vertex has a fitness $(a,b)$ with
probability $f(a,b)$. A vertex with fitness $(a,b)$, in-degree $i$ and
out-degree $j$ gains a new incoming edge with rate $a(i+1)$ and an outgoing
edge with rate $b(j+1)$. The Bose-Einstein condensation occurs as a function
of fitness distribution  $f(a,b)$.
\end{abstract}

\maketitle


\section{Introduction}

Recently there has been much interest in random growing networks \cite%
{review,review1,review2}, both from the point of view of theoretical
modelling, as well as the empirical study of real networks. There is
considerable evidence that while traditional Erd{\H o}s-Renyi random graphs have
a Poisson degree distribution \cite{bela}, real graphs and random growing
graphs have a power-law degree distribution.

Of particular interest are directed networks, which can be used to model
systems in which directed flow takes place. Such networks include the world
wide web \cite{1}, the phone-call graph \cite{2} and the citation graph \cite%
{3}.

As a complement to computer simulations and exact solutions of 
simplified systems, thermodynamic formulations have been used 
to study a number of complex co-operative systems 
including granular media, econophysics, breaking
phenomena and many others.
In a recent paper \cite{bb1}, Bianconi and Barab\'asi (BB) mapped the different
behaviour of random growing networks with fitness onto the 
thermodynamically distinct phases of a free
Bose gas. 
The
fitness model predicts that, in the large network limit, 
the fittest node will have the most
links. This is called the {\it fit-get-rich} (FGR) phase. Unlike the
scale-free model (SFM), in which the degree distribution of the 
network is power-law, 
the FGR behaviour (phase in the thermodynamic language) has
nodes with a very large degree which dominate the network. 
Another phase revealed in \cite{bb1} is the Bose-Einstein condensation
(BEC), when the fittest node acquires a finite fraction of the total degree. 
In contrast to BEC, in the FGR phase the richest node is not an
absolute winner, since its share of the links (\textit{i.e.} the ratio of
its degree and the total degree of the network) decays to zero for
large system sizes, whereas in BEC the winner maintains its share
irrespective of system size.

In this case, the fact that this ratio is constant corresponds to  
the extensivity property of a Bose gas, when the gas  
keeps a finite fraction of its particles in the
ground state. In this paper we examine the phenomenon of Bose-Einstein
condensation, which was considered for an undirected graph in \cite{bb,bb1}
on the random directed growing network introduced in \cite{er}.

To do this we work with the model introduced in \cite{er}, introducing a
dependence between the vertex fitness and an energy. As each vertex has two
fitnesses (for in-degree and out-degree) then it is necessary to assign two
different energy levels to represent each vertex. This is detailed in the
first sub-section, together with the description of edges in terms of
particles assigned to two energy levels.

The necessary coexistence of two sub-systems forming the network and the
formulation of an equilibrium condition in a formal way is discussed in
detail in sections 3 and 4.

The occurrence of Bose-Einstein condensation, and its interpretation in a
physical framework makes it possible to describe the directed network in
terms of canonical concepts of statistical physics, such as a phase diagram
and first and second order phase transitions. These are introduced in the
following sections.

In the conclusions we emphasize the usefulness of this formulation in
providing a clear and concise interpretation of different phenomena in
networks. So, the possibility of coexistence of phases in which the
in-degree distribution exhibits a clear winner whereas the out-degree shows
scale free behaviour, or the simultaneous existence of two different
winners (bipolarity) and the conditions for its existence are predicted in 
thermodynamic language.

\ \ \ \ \ \ \ 


\section{The Model}

In BB \cite{bb1}, a correspondence between fitness $\eta $ and energy $%
\epsilon $ given by 
\begin{equation}
\eta =e^{-\beta \varepsilon }
\end{equation}%
was introduced. In our model the vertices have, in general, different
fitnesses $a$ and $b$ for the addition of an in- or out-degree. As in \cite%
{er}, here the model consists of the addition of bare vertices (i.e. without
edges, but with fitness $a$ for in-degree and $b$ for out-degree) to the
network with probability $p$, and the creation of directed edges between
vertices with probability $q=1-p$.

\bigskip A kinetic equation describing the process of directed networks must
include the kinetics of both in-degree and out-degree simultaneously since
they coexist and influence each other. Then, the kinetic equation for the
mean number of vertices with in-degree $i$ and out-degree $j$ is, as in \cite%
{er},

\begin{equation}
\frac{\partial N_{ij}(a ,b )}{\partial t}=\frac{qa }{M_{1}}%
[iN_{i-1,j}-(i+1)N_{ij}]+\frac{qb }{M_{2}}[jN_{i,j-1}-(j+1)N_{ij}]+p\delta
_{i0}\delta _{j0}f(a ,b ).
\end{equation}%
where the fitnesses $(a,b)$ are chosen from the fitness distribution $f(a,b
) $. The normalisation factors $M_1$ and $M_2$ are given by 
\begin{equation}
M_{1}=\sum_{ijab }(i+1)a N_{ij}(a ,b )
\end{equation}%
and 
\begin{equation}
M_{2}=\sum_{ijab }(j+1)b N_{ij}(a ,b ).
\end{equation}%
The first term in the first square brackets of Eq.(2) represents the
increase in $N_{ij}$ when vertices with in-degree $i-1$ and out-degree $j$
gain an incoming edge and the second term represents the corresponding loss.
The second square brackets contain the analogous terms for outgoing edges
and the last term ensures the continuous addition of new vertices with
fitnesses $a ,b $.

The translation of this problem into an energetic formulation is
straightforward. We associate the addition of a vertex with fitnesses $(a,b
) $ to the network with the creation of two energy levels; one representing
the fitness of the incoming edge and the other for the outgoing edge. This
means that the edge can be mapped into two separate isolated sub-systems.
Then the creation of a directed edge corresponds to the creation of two
particles, one in each sub-system, simultaneously. The particles are created
in the energy levels corresponding to incoming fitness of the receiving
vertex and the outgoing fitness of the emitter vertex, as Fig. 1 illustrates.

Then, using the energy variables, rather that the fitness, we have 
\begin{equation}
a=e^{-\beta _{1}\varepsilon _{1}},
\end{equation}%
\begin{equation}
b=e^{-\beta _{2}\varepsilon _{2}}
\end{equation}%
and 
\begin{equation}
\frac{\partial N_{ij}(\varepsilon _{1},\varepsilon _{2})}{\partial t}=\frac{%
qe^{-\beta _{1}\varepsilon _{1}}}{M_{1}}[iN_{i-1,j}-(i+1)N_{ij}]+\frac{%
qe^{-\beta _{2}\varepsilon _{2}}}{M_{2}}[jN_{i,j-1}-(j+1)N_{ij}]+p\delta
_{i0}\delta _{j0}f(\varepsilon _{1},\varepsilon _{2}).
\end{equation}%
and the normalisations are re-written as 
\begin{equation}
M_{1}=\sum_{ijab}(i+1)e^{-\beta _{1}\varepsilon _{1}}N_{ij}(\varepsilon
_{1},\varepsilon _{2})
\end{equation}%
and 
\begin{equation}
M_{2}=\sum_{ijab}(j+1)e^{-\beta _{2}\varepsilon _{2}}N_{ij}(\varepsilon
_{1},\varepsilon _{2}).
\end{equation}

\bigskip As we distinguish between incoming and outgoing edges, and their
respective fitnesses, when going to the energy representation the edges and
their fitnesses must also be distinguished as belonging to different
isolated sub-systems of the same system. Let us denote the sub-systems as 1
and 2 \ So, the creation of a vertex in the network implies the simultaneous
creation of an energy level in each sub-system, i.e. $\varepsilon _{1\text{ }%
} $and $\varepsilon _{2\text{ }}$.

The creation of an edge joining two vertices implies the simultaneous
creation of one particle in each sub-system in the energy levels
corresponding to the fitnesses of the vertices gaining an in- and
out-degree, as Fig.1 illustrates.

It should be noted that we are doing a simultaneous description of two
isolated sub-systems which compose the whole system (the network). These
sub-systems do not exchange energy or particles, but are correlated in the
sense that the creation of an energy level (a particle) in one of the
sub-systems implies the simultaneous phenomenon in the other.

By definition of the model, the total number of particles increases linearly
with time, so 
\begin{equation}
\sum_{i,j,\varepsilon _{1},\varepsilon _{2}}N_{ij}(\varepsilon
_{1},\varepsilon _{2})=pt.
\end{equation}

Let us define the reduced moments $m_{1}$ and $m_{2}$ by 
\begin{equation}
M_{1}(t)=m_{1}t,
\end{equation}%
\begin{equation}
M_{2}(t)=m_{2}t
\end{equation}%
and introduce $n_{ij}$ as 
\begin{equation}
N_{ij}(t)=n_{ij}t.
\end{equation}

\bigskip As shall be seen these magnitudes will be useful to calculate the
characteristics of the network. Then, following the procedure in \cite{er},
we can show that
\begin{equation}
m_{1}=\frac{q}{m_{1}}\sum_{i,j,\varepsilon _{1},\varepsilon _{2}}e^{-2\beta
_{1}\varepsilon _{1}}(i+1)n_{ij}+p\int e^{-\beta _{1}\varepsilon
_{1}}f(\varepsilon _{1},\varepsilon _{2})d\varepsilon _{1}d\varepsilon _{2}
\end{equation}%
and 
\begin{equation}
m_{2}=\frac{q}{m_{2}}\sum_{i,j,\varepsilon _{1},\varepsilon _{2}}e^{-2\beta
_{2}\varepsilon _{2}}(j+1)n_{ij}+p\int e^{-\beta _{2}
\varepsilon_{2}}f(\varepsilon _{1},\varepsilon _{2})d\varepsilon _{1}d\varepsilon _{2}
\end{equation}
which determines $m_1$ and $m_2$ as functions of $p$, 
$\beta_1$ and $\beta_2$.

\subsection{The coexistence of two systems}

In this representation, as we already pointed out, the addition of a vertex
means the creation of two energy levels, each corresponding to its fitness.
When an edge joining two vertices is created, we add one particle in each
corresponding level.

The distribution of in- and out-degree then corresponds to the distribution
of particles in systems 1 and 2

\begin{equation}
g_{i}=\sum_{j}n_{ij}=p\frac{\Gamma (i+1)\Gamma (1+e^{\beta (\varepsilon
_{1}-\mu _{1})})}{\Gamma (i+2+e^{\beta (\varepsilon _{1}-\mu _{1})})}%
e^{\beta (\varepsilon _{1}-\mu _{1})}f(\varepsilon _{1},\varepsilon _{2})
\end{equation}

\bigskip is the number of vertices with in-degree $i$, and

\begin{equation}
h_{j}=\sum_{i}n_{ij}=p\frac{\Gamma (j+1)\Gamma (1+e^{\beta (\varepsilon
_{2}-\mu _{2})})}{\Gamma (j+2+e^{\beta (\varepsilon _{2}-\mu _{2})})}%
e^{\beta (\varepsilon _{2}-\mu _{2})}f(\varepsilon _{1},\varepsilon _{2})
\end{equation}

\bigskip is the number of vertices with out-degree $j$. For large degree
their asymptotic expressions are 
\begin{equation}
g_{i}\sim i^{-(1+e^{\beta (\varepsilon _{1}-\mu _{1})})}
\end{equation}%
and 
\begin{equation}
h_{j}\sim j^{-(1+e^{\beta (\varepsilon _{2}-\mu _{2})})}.
\end{equation}%
where we have introduced 
\begin{equation}
\mu _{k}=\frac{m_{k}}{q}
\end{equation}%
for $k=1$ and $k=2$ as \textit{chemical potentials}. Unlike BB, the chemical
potential is introduced here as an exact expression, not as an asymptotic
one.


\subsection{``Equilibrium" condition}

The equations for $m_{1}$ and $m_{2}$ can be transformed by introducing the
generating function 
\begin{equation}
g(x,y,\varepsilon _{1},\varepsilon
_{2})=\sum_{i,j}x^{i}y^{j}n_{ij}(\varepsilon _{1},\varepsilon _{2})
\end{equation}%
where 
\begin{equation}
g(1,1,\varepsilon _{1},\varepsilon _{2})=pf(\varepsilon _{1},\varepsilon
_{2})
\end{equation}%
with 
\begin{equation}
\frac{\partial g}{\partial x}(1,1,\varepsilon _{1},\varepsilon _{2})=\frac{%
pe^{-\beta _{1}\varepsilon _{1}}}{e^{-\beta _{1}\mu _{1}}[1-e^{-\beta
_{1}(\varepsilon _{1}-\mu _{1})}]}
\end{equation}%
and 
\begin{equation}
\frac{\partial g}{\partial y}(1,1,\varepsilon _{1},\varepsilon _{2})=\frac{%
pe^{-\beta _{2}\varepsilon _{2}}}{e^{-\beta _{2}\mu _{2}}[1-e^{-\beta
_{2}(\varepsilon _{2}-\mu _{2})}]}.
\end{equation}%
Then the equations for $m_{k}$ $(k=1,2)$ become 
\begin{equation}
\frac{p}{q}\int \frac{d\varepsilon _{1}d\varepsilon _{2}f(\varepsilon
_{1},\varepsilon _{2})}{e^{\beta _{k}(\varepsilon _{k}-\mu _{k})}-1}=1.
\end{equation}%
This means that 
\begin{equation}
\int \frac{d\varepsilon _{1}d\varepsilon _{2}f(\varepsilon _{1},\varepsilon
_{2})}{e^{\beta _{1}(\varepsilon _{1}-\mu _{1})}-1}=\int \frac{d\varepsilon
_{1}d\varepsilon _{2}f(\varepsilon _{1},\varepsilon _{2})}{e^{\beta
_{2}(\varepsilon _{2}-\mu _{2})}-1}.  \label{26}
\end{equation}%
Unlike BB, this is a new condition generated by the model. Let us denote it
as a \textit{generalized equilibrium condition}. A \emph{strong equilibrium}
condition implies that 
\begin{equation}
\varepsilon _{1}=\varepsilon _{2}
\end{equation}%
and 
\begin{equation}
\mu _{1}=\mu _{2}.
\end{equation}%
This condition will be discussed later. In this formalism the occupation
probability is, then, 
\begin{equation}
n_{k}(\varepsilon _{k},\mu _{k})=\frac{p}{q}\frac{1}{e^{\beta
_{k}(\varepsilon _{k}-\mu _{k})}-1},
\end{equation}%
for $k=1,2$, representing the probability for a given edge to belong to a
given vertex (probability of a particle to belong to an energy level). If,
now, the fitness distribution is separable then 
\begin{equation}
f(\varepsilon _{1},\varepsilon _{2})=f_{1}(\varepsilon
_{1})f_{2}(\varepsilon _{2})
\end{equation}%
and 
\begin{equation}
\frac{p}{q}\int \frac{d\varepsilon _{k}f(\varepsilon _{k})}{e^{\beta
_{k}(\varepsilon _{k}-\mu _{k})}-1}=1  \label{31}
\end{equation}%
separately for each system $k=1,2$. In each case, $\mu _{k}$ is the solution
of this equation, which depends on the density of states $f_{k}(\varepsilon
) $. 

\subsection{Bose-Einstein condensation}

To illustrate, let us start with a particular case, when $%
f(\varepsilon)=C\varepsilon ^{\theta }$ with $0 < \varepsilon <
\varepsilon_{\max}$. The normalisation $C=\frac{\theta +1}{\varepsilon
_{\max }^{\theta +1}}$ as in BB.

Then the integral in Eq.(25), 
\begin{equation}
I(\beta ,\mu )=\frac{p(\theta +1)}{q\varepsilon _{\max }^{\theta +1}}\int 
\frac{d\epsilon \epsilon ^{\theta }}{e^{\beta (\varepsilon -\mu )}-1}=1.
\end{equation}

But this integral reaches its maximum value when $\mu =0$. Then if for a
given value of $\beta $ we have 
\begin{equation}
\int \frac{d\varepsilon f(\varepsilon )}{e^{\beta \varepsilon }-1}<1,
\end{equation}
then we have a Bose-Einstein Condensation (BEC). If we calculate the
temperature for BEC in this particular case, we obtain 
\begin{equation}
T_{BE}=\frac{1}{\varepsilon _{\max }}\left\{ \frac{p}{q}\Gamma (\theta
+2)\zeta (\theta +1)\right\} ^{-\frac{1}{\theta +1}}.
\end{equation}

Unlike BB,the factor $\frac{p}{q}$ appears here. If $p=q$ we recover BB
result. Even in this simple example, $T_{BE}$ may be different for the
sub-systems 1 and 2, if the value of $\theta$ is different. In a more
general form, we can state the condition for Bose Einstein condensation.
Then, as 
\begin{equation}
I(\beta _{k},\mu _{k})=\frac{p}{q}\int \frac{d\varepsilon _{1}d\varepsilon
_{2}f(\varepsilon _{1},\varepsilon _{2})}{e^{\beta _{k}(\varepsilon _{k}-\mu
_{k})}-1}=1  \label{35}
\end{equation}%
for $k=1,2$., if $I(\beta _{k},\mu _{k})<1$ then BEC occurs. The conditions
for BEC depend on $f(\varepsilon _{1},\varepsilon _{2}).$


\subsection{Phase diagram}

Below $T_{BE}$ there is BEC. Above it the phase is fit-get-richer as in \cite%
{bb1}, but here both phases must be considered for both systems. Then, a
phase diagram can be drawn with the temperatures $T_{1}$ =$\frac{1}{\beta
_{1}}$ and $T_{2} $ =$\frac{1}{\beta _{2}}$ characterizing each sub-system
(Fig. 2).

Imposing a given dependence of fitness with energy so that 
\begin{equation}
\int \frac{d\varepsilon _{1}d\varepsilon _{2}f(\varepsilon _{1},\varepsilon
_{2})}{e^{\beta _{1}(\varepsilon _{1}-\mu _{1})}-1}\neq \int \frac{%
d\varepsilon _{1}d\varepsilon _{2}f(\varepsilon _{1},\varepsilon _{2})}{%
e^{\beta _{2}(\varepsilon _{2}-\mu _{2})}-1}
\end{equation}%
means that the equilibrium is broken, (i.e. in a computer simulation we would
impose temperatures that do not satisfy Eq. \ref{26}\ ). Then there can be a
winner in one of the phases. One of the vertices accumulates, say, most of
the in-degrees, coming from the other vertices. If both phases are in BEC
there is a vertex with most of the incoming and another with most of the
outgoing edges. Then both vertices will be highly correlated.


\subsection{Out of equilibrium first order phase transition}

Though a more complicated situations are possible, let us illustrate the
case $T_{1BE}=T_{2BE}$. The point A represents that both sub-systems are in
FGR phase. If, for example, we make the simulations with the parameters
given by A1 then will be an FGR and a BEC\ phase coexisting, but not \textit{%
in equilibrium} since one of them satisfies Eq. \ref{35} but not the other.
There is a coexistence of phases out of equilibrium. We might call this an 
\textit{out of equilibrium first order phase transition}. The motion along
AA1A2...leads to the region in which BEC occurs for both phases. Here is
hard to speak about equilibrium unless A belongs to the line $T_{1}=T_{2}$.
All we know in BEC region is that none of the sub-systems satisfies Eq. \ref%
{35}

The concept of equilibrium in this model is purely formal since the
sub-systems are not in contact, but it is a comfortable tool to make a map
of directed networks into Bose-Einstein statistics. 

\subsection{The scale free model}

If one of the systems is such that $f(\varepsilon )=\delta (\varepsilon )$,
then Eq. (\ref{31}) gives 
\begin{equation}
\frac{p}{q}\frac{1}{e^{-\beta \mu }-1}=1
\end{equation}%
then 
\begin{equation}
g_{i}\sim i^{-(1+e^{-\beta \mu })}\sim i^{-(2+\frac{p}{q})}
\end{equation}%
with a corresponding expression for $h_{j}$. If $p=q$ the scale free model
is recovered with a degree distribution $P_k \sim k^{-3}$. Thus the model
with directed edges gives the possibility of a scale free phase for both in
the in-degree and the out-degree. 

\subsection{Second order phase transition}

If in Fig. 3 we move in such a way as to cross into the BEC region through
the point B then BEC occurs simultaneously for both phases. No coexistence
of phases will occur, the system goes from FGR to BEC or vice-versa
instantly. This is a \textit{second order phase transition} since the system
as a whole changes its phase. Then the point B is the 
\textit{critical point} for the transition.

BEC is itself a second order phase transition since it effects the system as
a whole, so that when sub-system 1 or 2 experiences BEC, the sub-system
itself suffers a second order phase transition, but our system contains both
sub-systems 1 and 2 so that when they are in different phases we consider
our whole system as composed of two phases.


\section{Discussion}

We have studied Bose-Einstein condensation in a random growing network
model. In this model, with a formulation considering incoming and outgoing
edges, a formal analogy can be adopted to describe the change in the
behaviour of the directed edges in terms of Bose statistics.

In this formulation, as in that of BB, the temperature plays the role of a
dummy variable. Real systems are characterised by the functional form of the
distribution function of fitness and there is no $\beta$ to consider. It
only emerges when we translate from the language of fitness to that of
energies. But the formulation of networks in terms of quantum statistics and
the introduction of temperature is an elegant and simple description of the
behaviour of networks under different conditions. In this sense, the
temperature plays its role in the simulations of the networks, determining
the strength of the dependence between the fitness and its associated
energy, and this dependence is reflected in the behaviour of the network by
determining its position in the phase diagram.

Though phase diagrams and phase transitions are not exclusive to 
thermodynamic formulations (see, e.g. Dorogovtsev and Mendes \cite{dm}),
such a formulation is a very powerful tool to describe different phases of
random directed networks, where the presence of directed edges make
the description in terms of fitness difficult. Considering this network
as a thermodynamic system, makes it simple to understand, provided we 
appropriately
interpret the thermodynamic parameters.


\section{Acknowledgements}

We would like to thank The Royal Society, London and the London Mathematical
Society for their financial support. One of us (O. S) is greatly indebted to
the staff of the Department of Mathematical Sciences at Brunel 
University for their warm
hospitality. \newpage 

\section{Figure Captions}

1. Correspondence between fitness and energy for directed networks.

2. Phase diagram.

3. First and second order phase transitions. 

\bigskip


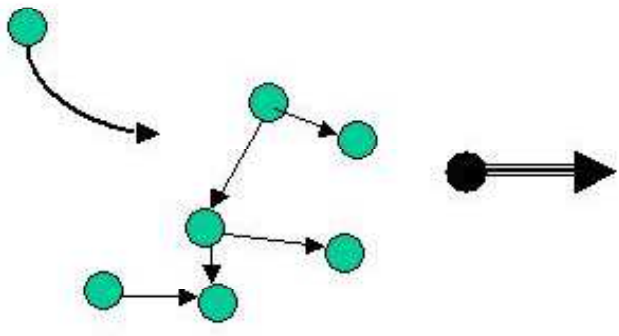
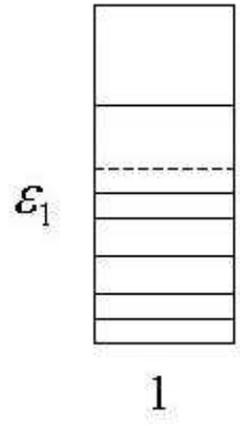
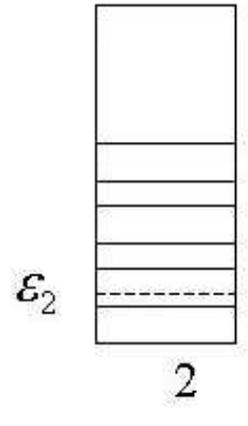
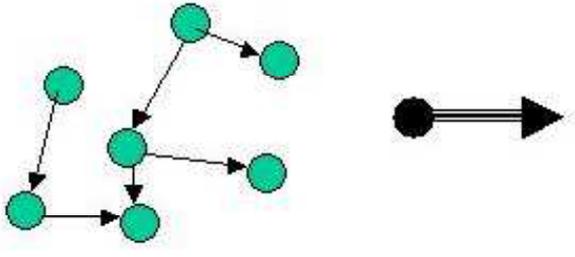
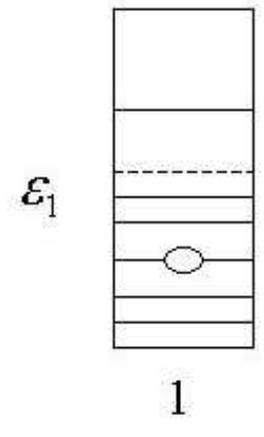
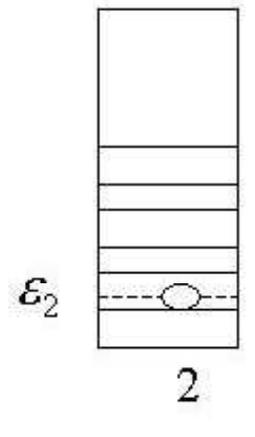
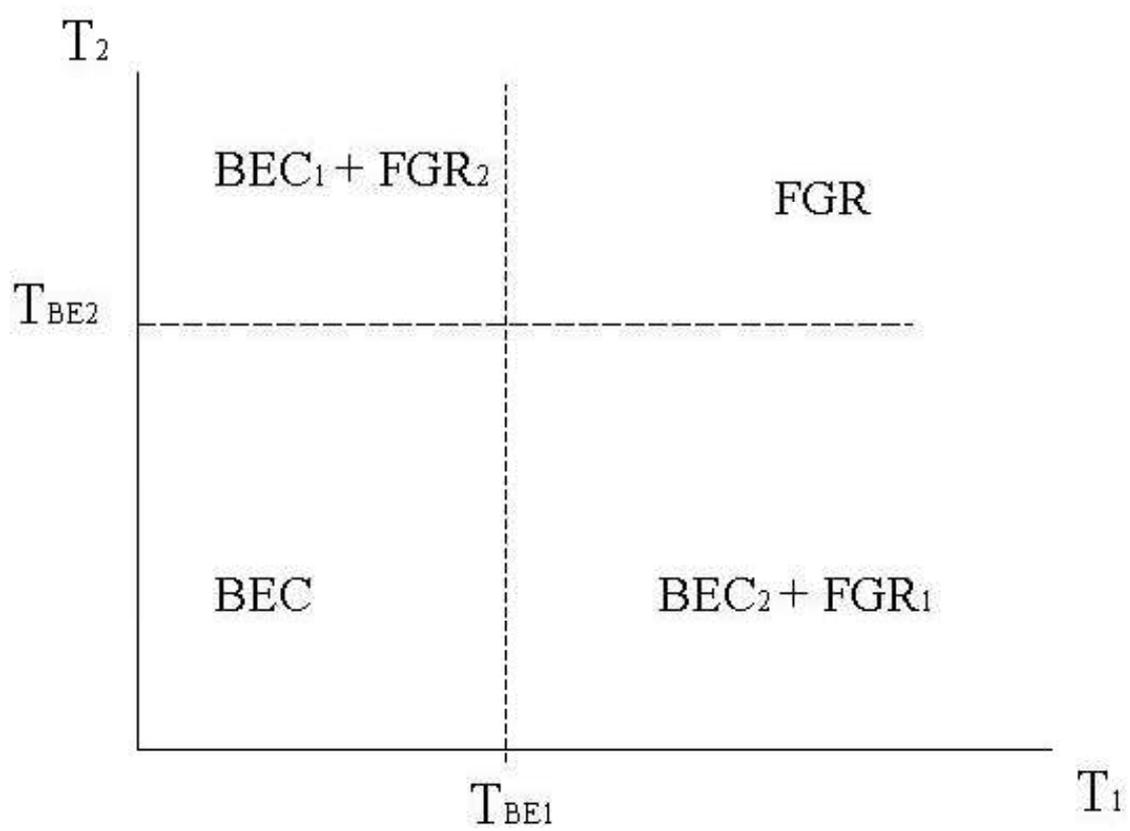
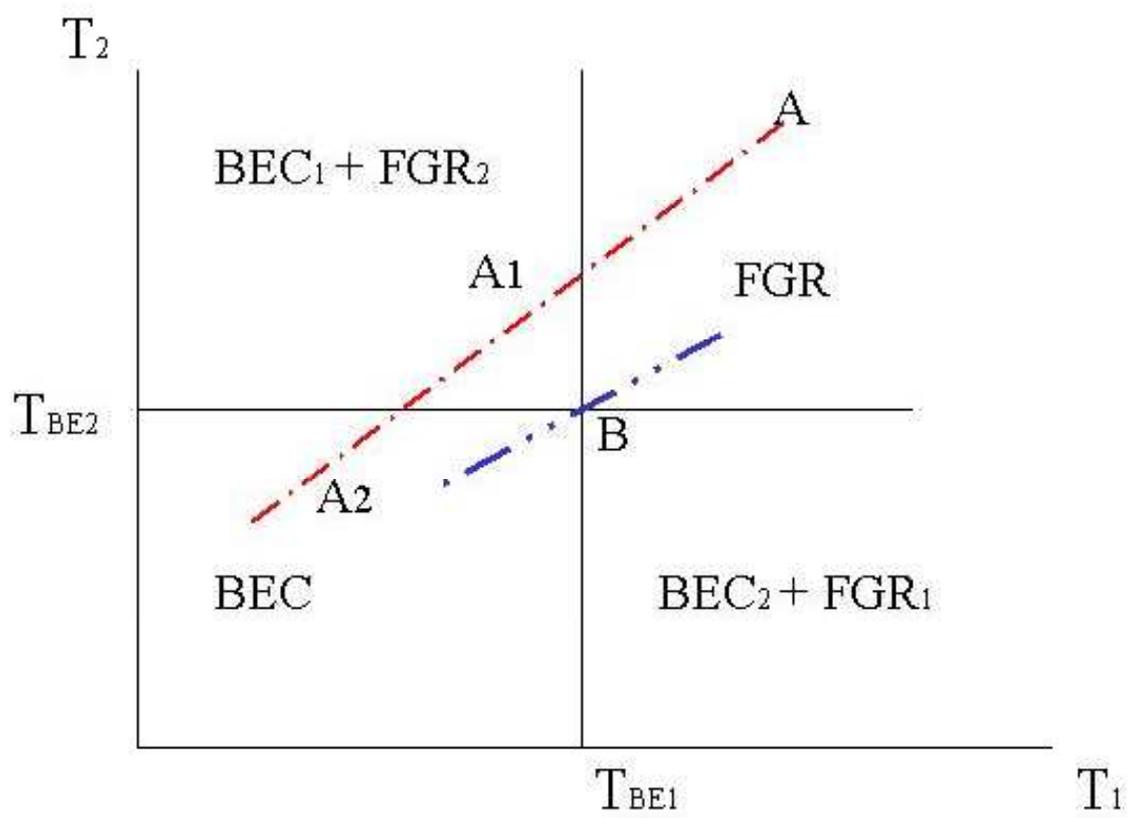
\end{document}